\documentclass{ws-procs11x85}
\usepackage{ws-procs-thm}           
\usepackage{algorithm}
\usepackage{algorithmic}
\usepackage{multirow}

\usepackage[table,xcdraw]{xcolor}
\usepackage{amsmath}
\usepackage{bbm}
\usepackage{algorithm}
\usepackage{algorithmic}
\usepackage{wrapfig}
\usepackage{pdfpages}
\usepackage{xcolor}

\begin{document}
\title{Scaling structural learning with NO-BEARS\\to infer causal transcriptome networks}

\author{Hao-Chih Lee$^{1,3}$, Matteo Danieletto$^{1,2,3}$, Riccardo Miotto$^{1,2,3}$,\\ Sarah T. Cherng$^{1,3}$ and Joel T. Dudley$^{1,2,3,\dag}$}

\address{
$^1$Institute for Next Generation Healthcare\\
$^2$Hasso Plattner Institute for Digital Health\\
$^3$Department of Genetics and Genomic Sciences\\
Icahn School of Medicine at Mount Sinai\\
New York, NY 10065, USA\\
$^\dag$E-mail: joel.dudley@mssm.edu}

\begin{abstract}
Constructing gene regulatory networks is a critical step in revealing disease mechanisms from transcriptomic data. In this work, we present NO-BEARS, a novel algorithm for estimating gene regulatory networks. The NO-BEARS algorithm is built on the basis of the NO-TEARS algorithm with two improvements. First, we propose a new constraint and its fast approximation to reduce the computational cost of the NO-TEARS algorithm. Next, we introduce a polynomial regression loss to handle non-linearity in gene expressions. Our implementation utilizes modern GPU computation that can decrease the time of hours-long CPU computation to seconds. Using synthetic data, we demonstrate improved performance, both in processing time and accuracy, on inferring gene regulatory networks from gene expression data.
\end{abstract}


\copyrightinfo{\copyright\ 2019 The Authors. Open Access chapter published by World Scientific Publishing Company and distributed under the terms of the Creative Commons Attribution Non-Commercial (CC BY-NC) 4.0 License.}

\section{Introduction}
Determining causal relations between attributes of observed data is a fundamental component of understanding biological systems. At a lower level, these systems typically consist of thousands of interacting attributes. While being observable individually, it is the intangible interactions among these attributes that provide mechanistic basis for the emergence of higher order biological functions. Using a transcriptome (i.e., a collection of genes expressed in an organism) as an example, a mutation in genetic codes can change interactions between genes and eventually manifest phenotypic disorders, such as cancer \cite{califano2017recurrent}. Quantifying the attributes and estimating the interactions are both important components of developing a deeper understanding of these systems. With recent advances in “omics” technologies, system-wide measurement of transcriptomic expressions is becoming a new paradigm within biomedical research. Efficient and accurate construction of biological networks from large-scale ‘‘omic” data is paramount for identifying the mechanisms of various interactions between attributes. 

Bayesian networks (BNs) is a methodological framework that can be used to infer probabilistic causal networks from biological data\cite{friedman2000using}. Combined with ‘’omic” technology, BNs have been used to predict genetic drivers of complex diseases including inflammatory diseases and Alzheimer’s disease\cite{zhang2013integrated,wang2012systems}. However, constructing a BN from observed data is computationally challenging, with computational complexity growing exponentially with the number of attributes measured\cite{chickering2004large}. This is typically achieved using score-based methods to sample structures from the discrete space of all possible networks that can lead to slow inference. Recently, NO-TEARS has been proposed as an alternative solution for estimating BNs by using continue-valued optimization\cite{zheng2018dags}. This method utilizes a novel constraint that enables solving structural learning problems via an efficient optimization method. Despite being non-convex optimization, Zheng et al. demonstrate the possibility of the algorithm converging toward the global solution. However, evaluating the constraint used in the NO-TEARS algorithm requires numerical operations that grows as a cubic function of the number of attributes, imposing a computational bottleneck for practical use.

This work aims to improve the NO-TEARS algorithm for a fast construction of BNs from transcriptomic data using principled optimization methods. Our contributions include:
\begin{itemize}
\item Introducing a new constraint that is theoretically equivalent to the one used in the NO-TEARS algorithm, and can be approximately evaluated in $O(n^2)$ steps.
\item Proposing a new regression loss to handle non-linearity in gene expression data
\item Combining the proposed constraint and the regression loss to develop a new algorithm, named NO-BEARS, and demonstrate its superior performance in both processing time and accuracy.
\item Accelerating computation using a graphical processing unit (GPU) that can substantially decrease processing time from hours to seconds
\end{itemize}
The implementation is available at \url{https://github.com/howchihlee/BNGPU}.
\section{Background}
\subsection{Gene regulatory network and the linear structural equation model}
The expression of genes are tightly regulated to maintain biological functions. The regulatory mechanism can be modeled as\cite{liao2003network}

\begin{equation}
x_i = \sum_{j \in pa(x_i)}W_{ij} x_j + \epsilon.
\label{eq:sem}
\end{equation}

This model, derived from the Hill equation for chemical reactions, assumes that the expression of a gene $x_i$ is jointly regulated by its parent genes $x_j \in pa(x_i)$ that are ``upstream" in the pathway. The regulatory effects are assumed to be linearly additive where $W_{ij}$ specifies the size of the effect exerted from gene $j$ to gene $i$. 
Overall, the gene regulatory network can be seen as a weighted directed graph, represented by $W$. In this directed graph, a node is a gene and an edge from $x_j$ to $x_i$ encodes the regulatory effect from gene $j$ to gene $i$. In the broader context of causal relation inference, this model is also called the {\it general linear structural equation model}. 
\vspace{-2mm}
\subsection{NO-TEARS Algorithm}
Determining how best to measure $W$ from observed values $x_i$ is the key to identifying gene regulatory networks. In many cases, directed acylic graphes (DAGs) are utilized to represent the causal system for its interpretability. The NO-TEARS algorithm is a new approach to identify the causal DAG using a continuous-valued objective function. At the core of the NO-TEARS algorithm is the matrix exponential $\exp(A)$, which measures whether a non-negative matrix $A$ represents a DAG. Specifically, Zheng et al. show that, the condition $tr(\exp(A)) = d$, where $tr(\cdot)$ is the trace operation, is equivalent to that the graph represented by the $d\times d$ adjacency matrix $A$ is acyclic. This is deducted from the fact that $tr(A^k) = 0$ is equivalent to there being no $k$-cycle in the adjacency matrix $A$. Given that there is no $k$-cycle for all $k\geq 1$ in a DAG, one can conclude that
\[tr(\exp(A)) = \sum_{k=0}^{\infty} \frac{1}{k!}tr(A^k) = tr (I_d) + \sum_{k=1}^{\infty} \frac{1}{k!}tr(A^k) = tr (I_d) = d \]

Zheng et al. thus propose to solve a constrained optimization problem to learn the causal structure: given an observed data matrix $X\in \mathbb{R}^{n\times d}$ of $n$ $d$-dimensional samples, find $W\in \mathbbm{R}^{d\times d}$ that solves 
\[\begin{array}{ll}
\min_{W\in R^{d\times d}} &  \frac{1}{2n}\|X - XW\|_F^2  + \beta\|W\|_1 \\
\textup{subject to } & tr(\exp(W\odot W)) = d
\end{array}
\] 
This optimization problem can be interpreted as follows: among all possible weighted DAGs, we want to identify one that is most consistent with the equation (\ref{eq:sem}). The $l_1$ loss is used to regularize the learning algorithm when the sample size is small. One drawback of the NO-TEARS algorithm is its computational complexity. Evaluating a matrix exponential requires $\mathcal{O}(d^3)$ numerical operations, making this optimization problem computationally expensive to solve. Another is that the regression loss assumes linearity in gene expressions, which may not be true. We seek to address these two issues to improve the NO-TEARS algorithm.   

\section{Original contribution}
In Section \ref{sec:DAG}, we present an alternative constraint, the spectral radius of a matrix, that also enforces directed acyclic properties, similar to the matrix exponential. The advantage of using this alternative penalty term is that it can be approximated with $\mathcal{O}(d^2)$ computational complexity, compared to the $\mathcal{O}(d^3)$ method of evaluating the matrix exponential used in the NO-TEARS algorithm. Importantly, the approximation is sufficiently scalable to infer a $d\times d$ gene regulatory network. In addition, we propose a new regression term that incorporates polynomial regression to address potential non-linearity of the dynamics of gene expression (Section {\ref{sec:poly_reg}}). Putting these two terms together, we propose a new optimization problem for learning causal networks from data (Section \ref{sec:opt_problem}).
\vspace{-2mm}
\subsection{Alternative characterization of DAG}
\label{sec:DAG}
To justify the proposed characterization for DAGs, we first demonstrate that the condition of no cycles for a non-negative matrix is equivalent to that its eigenvalues are all zero. The mathematical proof of this assertion can be found in the appendix. 
\begin{theorem}
Given $A\in \mathbb{R}^{d\times d}$ a real square matrix, the following statements are equivalent:
\begin{enumerate}
\item $tr(A^k) = 0$ for $k = 1,\dots d$, where $tr(\cdot)$ is the trace of a matrix.
\item $\sigma(A) = \{0\}$, where $\sigma(A)$ is the set of eigenvalues of $A$.\\
\vspace{-4mm}
\end{enumerate}
\end{theorem}
 
The implication of this theorem is that one can impose the directed acyclic property by restricting the solution space to be matrices with zero eigenvalues. We thus propose to use the spectral radius $\rho(A)$ to measure whether the graph $A$ has cycles or not. The spectral radius of a matrix is defined as the maximum of the absolute values among all eigenvalues of a matrix. Although evaluating the spectral radius of a $d\times d$ matrix may appear to require $\mathcal{O}(d^3)$ operations inherited from eigenvalue decomposition, we later present a way to approximate the spectral radius with $\mathcal{O}(d^2)$ operations.
\vspace{-2mm}
\subsection{Polynomial regression loss}
\label{sec:poly_reg}
The dynamics of gene expression are intrinsically nonlinear such that the equation (\ref{eq:sem}) may be a poor approximation. To handle the nonlinearity associated with gene expression, we instead extend the equation (\ref{eq:sem}) to
\[X_i = \sum_{j\in pa(i)}W_{ij}f_j(X_j) + \epsilon,\]
where $f_j$ is a scalar function that maps a gene $X_j$ into a space where equation (\ref{eq:sem}) holds true. The function $f_j$ is assumed to be not dependent on the response $x_i$ for the purpose of reducing the number of parameters. In practice, we parameterize $f_j$ to be a polynomial of degree $K$, where $K$ is set to be 3 in all of our experiments. This leads to a regression term, which we call the polynomial regression loss, defined as
\[PR(X, W, \alpha) = \sum_{il}\left(X_{il} - \sum_{j\in pa(i)}{W_{ij}} \sum_{k=0}^{K}\alpha_{jk}(X_{jl})^k\right)^2. \] 
\vspace{-4mm}
\subsection{Proposed optimization problem}
\label{sec:opt_problem}
Combining the spectral radius and the polynomial regression loss, we propose an alternative optimization problem for structural learning: Given a data matrix $X \in \mathbb{R}^{n\times d}$ representing $n$ samples of $d$ measured gene expressions, we estimate the causal network of $d$ genes by solving $W^*$ that minimize the following problem
\[\begin{array}{ll}
\min_{W\in R^{d\times d}, \; \alpha \in \mathbbm{R}^{d\times K}} &  \frac{1}{2n} PR(X, W, \alpha) + L(W, \alpha) \\
\textup{subject to } & \rho(W\odot W) = 0
\end{array}\]
where 
$L(W, \alpha) = \beta_1(\|W\|_F^2 + \|W\|_1) +\beta_2 \|\alpha\|^2_F$ is the regularization loss controlling model complexity and $\odot$ is the Hadamard product.

\section{NO-BEARS Algorithm and approximations}
We use the augmented Lagrangian method to solve the optimization problem. The augmented Lagrangian method consists of two general steps:
\begin{enumerate}
\item solve the unconstrained problem 
\begin{equation}
W,\; \alpha \leftarrow \textup{argmin}_{W, \alpha} \frac{1}{2n}PR(X, W, \alpha) + L(W, \alpha) + \xi \rho(W\odot W) + \frac{\eta}{2} \rho(W\odot W)^2;
\label{equation:main_problem}
\end{equation}

\item update the Lagrange multiplier $\xi \leftarrow \xi + \eta \rho(W \odot W)$; optionally one can increase $\eta$ for faster convergence. 
\end{enumerate}
In principle, any gradient-based optimization algorithm can be used to iteratively solve the unconstrained problem. The major challenge lies in whether we can efficiently evaluate the gradient of $\rho(W\odot W)$ with respect to $W$. We next present our method, named the NO-BEARS algorithm, for estimating the gene causal network. The NO-BEARS algorithm is summarized in Algorithm \ref{alg:inner} and \ref{alg:nobears}.

\begin{algorithm}
\caption{NO-BEARS inner iteration}
\label{alg:inner}
\begin{algorithmic}
\STATE input $W$, $\alpha$, $u\geq 0$ and $v\geq 0$, $\eta$
\STATE  solve the problem in equation (\ref{equation:main_problem}) by
\REPEAT
\STATE compute $\hat{W} = W\odot W + \epsilon$. Use burn-in without re-initializing $u$ and $v$. 
\STATE compute $u \leftarrow \frac{\hat{W}^Tu}{\|\hat{W}^Tu\|_2}$ and $v \leftarrow \frac{\hat{W} v}{\|\hat{W} v\|_2}$ for 5 iterations.
\STATE obtain $W^+$, $\alpha^+$ using any gradient based algorithm to minimize
\[\frac{1}{2n} PR(X, W, \alpha) + L(W, \alpha) + \xi  \frac{u^T\hat{W}v}{u^Tv} + \frac{\eta}{2} (\frac{u^T\hat{W}v}{u^Tv})^2\] by one step. Assign $W \leftarrow W^+$, $\alpha \leftarrow \alpha ^+$
\UNTIL{converged or a maximum number of iterations is reached} 
\RETURN $W, \alpha, u, v$
\end{algorithmic}
\end{algorithm}
\vspace{-6mm}
\begin{algorithm}
\caption{NO-BEARS algorithm}
\label{alg:nobears}
\begin{algorithmic}
\STATE initialize $W$, $\alpha$, $u\geq 0$ and $v\geq 0$, $\eta$ 
\REPEAT
\STATE $W, \alpha, u, v \leftarrow$ NO-BEARS inner iteration($W,\alpha,u, v,\eta$)
\STATE update the Lagrange multiplier $\xi \leftarrow \xi + \eta \rho(W \odot W)$, $\eta \leftarrow \eta \times 1.1$ 
\UNTIL{converged or a maximum number of iterations is reached}
\RETURN $W$
\end{algorithmic}
\end{algorithm}
\vspace{-6mm}
\subsection{Make NO-BEARS differentiable}
\label{sec:nobear_diff}
To solve an optimization problem using a gradient based algorithm, we must first determine whether the objective function is differentiable. Unfortunately, the spectral radius $\rho(W)$, as a function of a matrix $W$, is not necessarily differentiable. To resolve this problem, we instead minimize a perturbed objective function whose gradient can be evaluated. Specifically, we propose to solve
\[W^*,\; \alpha^* = \textup{argmin}_{W, \alpha} \frac{1}{2n}PR(X, W, \alpha) + L(W, \alpha) + \xi_k \rho(W\odot W + \epsilon) + \frac{\eta}{2} \rho(W\odot W + \epsilon)^2\]
where $\epsilon$ is a small constant, chosen to be $10^{-6}$ in our implementation, added to all elements of the matrix $W$. The approximated problem provides an upper bound to the original problem since $\rho(W\odot W)\leq\rho(W\odot W + \epsilon)$ is always guaranteed by the Wielandt's Theorem \cite{meyer2000matrix}. In the appendix we show that $\rho(W\odot W + \epsilon)$ is differentiable with a gradient
\[\frac{\partial \rho(W\odot W + \epsilon)}{\partial W_{i,j}} = 2\frac{W_{ij}v_iu_j}{u^Tv},\] where $u$ and $v$ are left and right eigenvectors of $W\odot W + \epsilon$. Any gradient descent algorithm can then be used to solve the proposed optimization problem, assuming that $u$ and $v$ can be computed accordingly. The question that then naturally arises is determining whether the eigenvectors can be estimated efficiently.
\vspace{-4mm}
\subsection{Make NO-BEARS scalable}
We describe how to efficiently compute the right eigenvector of $W\odot W + \epsilon$ associated with its largest eigenvalue by using power iteration. The left eigenvector can be computed in a similar way. Power iteration is an iterative method that computes the eigenvector associated with the largest eigenvalue of a matrix. Starting from a vector $v_0\in \mathbb{R}^{d}$, it recursively computes 
\[v_{k+1} = \frac{(W\odot W + \epsilon) v_k}{\|(W\odot W + \epsilon) v_k\|_2}.\] Since $W\odot W + \epsilon$ has a dominant eigenvalue that is strictly larger than other eigenvalues in absolute value, the sequence $v_k$ is guaranteed to converge to the eigenvector of the dominant eigenvalue, despite presenting with a slow convergence rate. To further accelerate convergence, we adapt the idea of spectral normalization, a technique recently propose to regularize deep neural networks\cite{miyato2018spectral}. Specifically, we use the estimated eigenvector $v$ from the previous iteration as a start from burn-in, and refine it using power iteration (see Algorithm \ref{alg:inner}). In practice we find that a few iterations is sufficient to obtain the eigenvector with  reasonable accuracy using this burn-in strategy. Since this method requires only a fixed number of multiplications of the matrix $W$ by a vector $v$, the computational complexity to approximate $\rho(W\odot W + \epsilon)$ and its gradient is $\mathcal{O}(d^2)$ for any $d\times d$ matrix $W$. 

\section{Implementation details of NO-BEARS algorithm}
We implement the proposed algorithm using tensorflow\cite{abadi2016tensorflow}. All of our experiments are conducted on 32-core server with a NVIDIA TITAN X Graphics card. In the following, we discuss a few techniques that empirically improve the performance of the proposed NO-BEARS algorithm.
\vspace{-2mm} 
\subsection{Initialization}
We initialize the weight matrix $W$ by fitting the equation (\ref{eq:sem}) without the acyclic constraint. While fitting the equation (\ref{eq:sem}) requires utilizing the unknown set of parent genes, we instead estimate the multiple linear dependencies of the gene $X_i$ on a set of highly {\it correlated} genes. Specifically, we first select up to 30 genes based the absolute values of their Spearman correlation with respect to $X_i$. The edge weight $W_{ij}$ of these genes is then estimated using multiple linear regression while $W_{ij}$ is set to be 0 if the gene $j$ is not selected.
We further refine $W$ by solving a minimization problem
\begin{equation}
W^+,\; \alpha^+ = \textup{argmin}_{W, \alpha} \frac{1}{2n}PR(X, W, \alpha) + L(W, \alpha)
\label{equation:initialization}
\end{equation}
using a first-order gradient descent optimizer. We used the ADAM optimizer\cite{kingma2014adam} for all of our experiments to iteratively update $W$ and $\alpha$ for 200 steps.

\vspace{-2mm}
\subsection{Data augmentation}
Augmenting data is known to facilitate the training of deep neural networks. Here, we discuss two strategies for data augmentation. The first method involves adding random noise to input data at every iteration. We find that adding $10\%$ Gaussian noise, model training generally improves. We adapt the idea of bootstrapping, a technique for reducing variance in statistical estimation, as well. Specifically, at every iteration, we shuffle the data while allowing the same sample to be drawn twice, at most. To create a GPU-friendly implementation, we first replicate the data matrix two-fold, reshuffle the order, and then perform a mini-batch iteration of batch size that is the same as the sample size. We empirically find that this strategy performs better than sampling with replacement.
\vspace{-2mm}
\subsection{Moving averaging}
Weight-averaged solutions can sometimes contribute to more accurate models\cite{tarvainen2017mean}. We experiment with this technique and find that calculating the moving average of weight matrix $W$ over the outer iterations indeed results in improved model performance. For our experiments, we specifically chose the moving average rule to be 
$W_{t+1} = 0.975 \times W_t + 0.025 \times W^+$,
where $W^+$ is the output from Algorithm \ref{alg:inner}. The last $W_t$ is outputted as the estimated weight matrix.
\vspace{-2mm}
\subsection{Stopping criteria}
We design the stopping criteria based on an empirical observations: the best performing solution is typically attained after the regression loss rebounds from its minimum. Generally speaking, a good solution requires balancing the regression loss that enforces data faithfulness and the spectral-radius penalty that enforces acylicness. While a low spectral-radius penalty encourages acylicness, too much acylicness would eventually override data faithfulness. We thus stop the NO-BEARS algorithm when 1) the spectral radius is smaller than a threshold $t_0$ and 2)
the regression loss is $t_1$-fold larger the observed minimum over iterations. We choose $t_0$ and $t_1$ to be $0.005$ and $1.5$ for all our experiments.

\section{Evaluation}
We compare NO-BEARS with two baseline models, NO-TEARS (section 6.1) and GENIE3 (section 6.2). We evaluate the performance using two sets of synetic data simulated by  Synthetic Transcriptional Reg-ulatory Networks (Section 6.3) and GeneNetWeaver (Section 6.4).
\subsection{Baseline model: NO-TEARS algorithm using GPU}
We implement the NO-TEARS algorithm using tensorflow as follows:
\begin{enumerate}
\item initialize $W$
\item ({\bf inner iteration}) use ADAM algorithm to minimize  
\begin{equation}
\frac{1}{2n}\|X - XW\|_F^2 + \beta\|W\|_1 + \xi h(W) + \eta h(W)^2,
\label{equation:no_tears}
\end{equation} 
where $h(W) = tr\exp(W\odot W) - d$. Iterate until converge or a maximum number of iterations is reached. We iterate the solution for 100 steps. 
\item update the Lagrange multiplier $\xi^+ = \xi + \eta h(W)$ and $\eta^+ = 1.25 \times \eta$ 
\item repeat 2 and 3 until converge ($h(W)$ is less than $10^{-8}$) or a maximum number of iterations, which we choose to be 50, is reached.
\end{enumerate}
We re-implemented the NO-TEARS algorithm to evaluate and benchmark its performance. The original implementation uses only CPU computation that results in the algorithm being too slow to compare with other methods. For example, the original implementation, using 32 cores in parallel, takes $\sim 4000$ seconds to process simulated data with 300 genes while our GPU implementation of the algorithm only requires $\sim 40$ seconds. 

\vspace{-4mm}
\subsection{Baseline model: GENIE3}
We use another baseline model, GENIE3, which is the top performing model in DREAM4 challenges\cite{irrthum2010inferring} as another benchmarking method. GENIE3 is designed based on a linear structural equation model. Instead of using linear regression to fit the structure model, GENIE3 utilizes Random Forest regression and ranks the gene-gene dependency by feature importance. We use the original Python implementation with recommended parameters in our experiments. GENIE3 is benchmarked using parallel processing on 20 CPUs.

\begin{table}[h]
\tbl{Characteristics of benchmark datasets}{
\fontsize{10}{10}\selectfont
\begin{tabular}{lcccccccc}
\toprule
&DAG-100 & DAG-300 & Ecoli-100 & Ecoli-300 & Yeast-100 & Yeast-300 & Ecoli-1565 & Yeast-4441\\
\colrule
$\#$nodes &100 & 300 &100 & 300 &100 & 300 & 1565 & 4441\\
$\#$edges &121 & 427 & 164& 463 & 149 & 543 & 3648 & 12873\\
$\#$parents & $1.2 \pm 4.3$ & $1.4 \pm 5.4$ & $1.6 \pm 5.3$ & $1.5 \pm 5.6$ & $1.5 \pm 3.0$ & $1.8 \pm 4.7$ & $2.3 \pm 18.2 $& $2.9 \pm 21.8$ \\
\colrule
$\#$samples &  \multicolumn{6}{c}{100, 500, 1000 and 2000} & 1565 & 4441\\
\colrule
simulator &  \multicolumn{6}{c}{SynTReN} & \multicolumn{2}{c}{GeneNetWeaver}\\
\botrule
\end{tabular}\label{table:datasets}
}
\vspace{-8mm}
\end{table}
\subsection{Benchmark datasets}
To benchmark performance, we use two sets of synthetic data that are simulated based on curated transcriptome networks. The first set is generated by Synthetic Transcriptional Regulatory Networks (SynTReN \cite{van2006syntren}). We use SynTReN to simulate expression data of 100 and 300 genes. The sample size of each simulation ranges from 100, 500, 1000 and 2000. Three different types of ground truth graphs, including a directed acylic graph, an E. coli gene network and a S. cerevisiae gene network, are used to simulate data. The second set of data is simulated using GeneNetWeaver\cite{schaffter2011genenetweaver}. We use a data set generated in a previous study\cite{statnikov2015ultra}. This dataset includes data simulated based on E. coli and S. cerevisiae gene networks, each with 1565 and 4441 genes, respectively. Table \ref{table:datasets} summarizes characteristics of all data that we use for data simulation.
\vspace{-2mm}
\subsection{Metrics}
We treat the gene-network inference as a binary prediction problem. The task is to predict the existence of an gene-gene edge using gene expression data. We therefore measure the accuracy of inferring gene-gene dependency using the area under the receiver operating characteristic curves (AUC-ROC) and the average precision score (AP) by using the absolute values of outputted weights $W$ as scores to predict gene-gene edges.
\section{Result}
\subsection{Benchmarks on SynTeRN simulations}
\begin{figure}
\centering
\centerline{
\includegraphics[width=\textwidth]{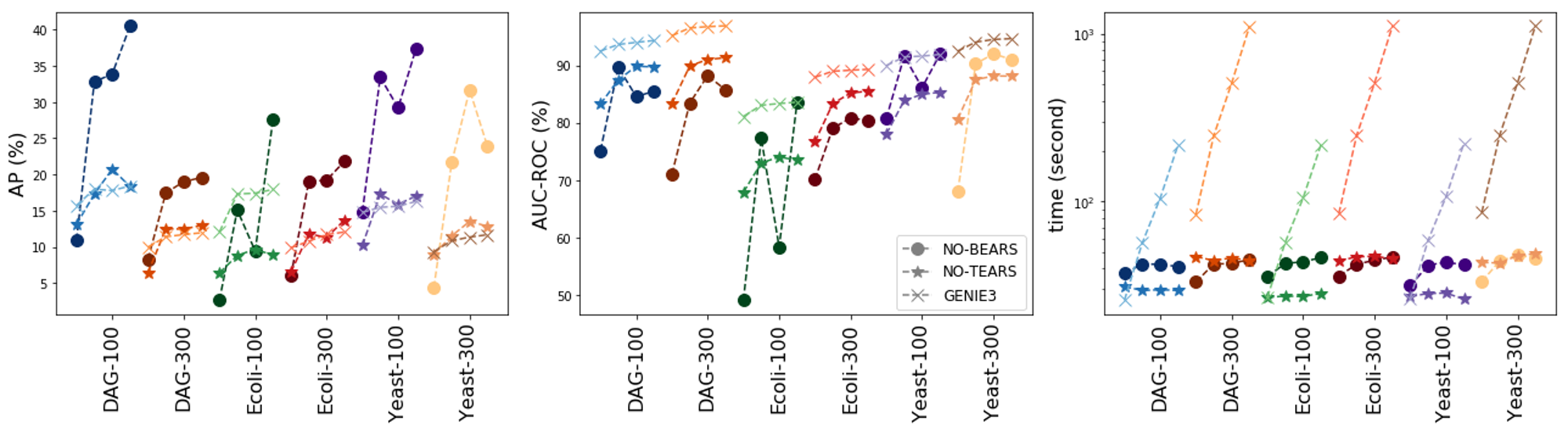}}
\caption{Performance benchmark on SynTReN. From left to right, data points of the same color are arranged with sample sizes increasing from 100, 500, 1000 to 2000. Circles, stars and cross marks indicate results obtained using NO-BEARS, NO-TEARS and GENIE3 respectively. Panels from left to right show average precision scores (AP), area under a ROC curve (AUC-ROC) and processing time.}
\vspace{-7mm}
\label{fig:syntern_benchmark}
\end{figure}

Benchmarks on data simulated by SynTeRN are reported in Figure \ref{fig:syntern_benchmark}. In general, NO-BEARS outperforms the other two methods by a significant margin in AP scores when the sample size of edges is larger than 500. Despite observing a drop in AUC-ROC, we note that the edge classification is highly imbalanced, with no more than 10\% positive gene-gene dependencies over the combination of all possible edges which can bias AUC-ROC. For all cases tested, NO-BEARS and NO-TEARS are able to return results in 1 minute while GENIE3 took $\sim$20 minutes to process cases of 2000 samples. We note that processing time measures the total convergence time, and is not expected to be square for NO-BEARS nor cubic for NO-TEARS.

\vspace{-5mm}
\subsection{Benchmarks on GeneNetWeaver simulations}
\begin{wraptable}{r}{0.55\linewidth}
\vspace{-4mm}
\tbl{Benchmark on data sets simulated by GeneNetWeaver. AP: average precision (\%). ROC: Area under a ROC curve (\%). time: processing time (s).}{\fontsize{10}{9}\selectfont
\begin{tabular}{llccc}
\toprule
                            &        & NO-BEARS & NO-TEARS & GENIE3  \\
\colrule
\multirow{3}{*}{Ecoli-1565} & AP   & 36.1     & 35.2     & 12.2    \\
                            & ROC & 78.7     & 83.0     & 88.0    \\
                            & time   & 154.9    & 352.5    & 5538.5  \\
\colrule
\multirow{3}{*}{Yeast-4441} & AP   & 64.5     & 56.9     & 9.8     \\
                            & ROC & 95.6     & 88.9     & 92.9    \\
                            & time   & 2484.1   & 4842.8   & 89431.2 \\
\botrule
\end{tabular}
}

\label{table:gnw_benchmark}
\vspace{-6mm}
\end{wraptable}
Benchmarks on data simulated by GeneNetWeaver are reported in Table \ref{table:gnw_benchmark}. In these two cases, NO-BEARS achieved the highest average precision scores using least amount of time. The significantly shorter processing time for NO-TEARS and NO-BEARS suggests the advantage of using GPUs for solving computationally challenging problems.
\vspace{-4mm}
\subsection{Effects of minimizing spectral radius}
\begin{figure}[ht]
\vspace{-6mm}
  \begin{minipage}[c]{0.6\textwidth}
    \includegraphics[width=0.95\textwidth]{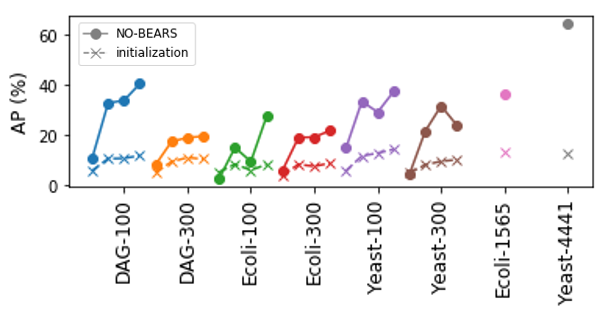}
  \end{minipage}\hfill
  \begin{minipage}[c]{0.4\textwidth}
    \caption{
       Comparing the performance with and without minimizing the spectral radius. Circles and cross marks indicate average precision scores obtained with and without applying spectral radius minimization respectively. From left to right, data points of the same color are arranged with sample sizes increasing from 100, 500, 1000 and 2000.
    } \label{fig:compare_initializer}
  \end{minipage}
 \vspace{-6mm}
\end{figure}

We also investigate whether imposing structural constraint improves performance. We compare initial solutions that are computed as described in section 4.1 to final solutions that are further refined by minimizing the full objective function (equation \ref{equation:main_problem}) until stopping criterion are met. Figure \ref{fig:compare_initializer} shows the AP scores of these initial solutions and their corresponding final solutions. As it can be seen, enforcing directed acycliness improves estimating gene networks.

\subsection{Computational complexity}
Last, to benchmark computational complexity of our problem, we measure the time to process 100 inner iterations for the NO-BEARS algorithm (equation \ref{equation:main_problem}) and NO-TEARS (equation \ref{equation:no_tears}). The results are reported in table \ref{table:complexity}. We note that these results only reflect the processing time of a fixed number of iterations, but not the overall time till converge to a solution.
\begin{table}[h]
\vspace{-2mm}
\tbl{real-world time to process 100 iterations. All numbers are reported in seconds. OOM: Out-of-memory associated results are not available.}{
\begin{tabular}{lccccccccc}
\toprule
$\#$genes         & 100  & 200  & 400  & 800  & 1600 & 3200  & 6400   & 12800 & 25600 \\
\colrule
NO-BEARS & 0.96 & 0.70 & 0.74 & 0.73 & 0.79 & 1.14  & 3.14   & 9.70  & OOM   \\
NO-TEARS & 1.05 & 1.04 & 1.56 & 2.98 & 9.00 & 36.97 & 241.10 & OOM   &   \\   
\botrule
\end{tabular}
}
\label{table:complexity}
\vspace{-10mm}
\end{table}
\section{Discussion and conclusion}
We present the NO-BEARS algorithm for estimating gene regulatory networks. the NO-BEARS algorithms is built on the basis of the NO-TEARS algorithm with a new penalty term to improve scalability, and the application of a new regression loss function to combat non-linearity in gene expression dynamics. We observe a faster and more accurate construction of gene regulatory networks on several synthetic data sets using the NO-BEARS algorithm. Despite these promising results, further evaluation is required to test the algorithm's performance on real-world data. We observed a sample-size dependency on the performance of BN construction in three tested methods (Figure 1), suggesting that sample size is a critical factor to consider when deploying BN construction in a real-world setting. In this work, we demonstrate that GPUs can greatly accelerate BN construction, despite a trade off with GPU memory, which limits the size of the BN. The NO-BEARS algorithm is memory efficient, compared to the NO-TEARS algorithm, and can handle up to 12,800 genes using 1 GPU, a capability that is applicable to most real world scenarios. As GPU hardware improves, we believe constructing a full transcriptome network is possible in the near future. In addition to rapidly evolving sequencing techniques to probe the molecular makeup of an individual, we envision that this method will contribute to a better understanding on the role of gene regulatory networks in personalized health.
\section*{Acknowledgments}
The authors would like to thank support from the Hasso Plattner Foundation and a courtesy GPU donation from NVIDIA.
\vspace{-4mm}
\section{Appendix}
\subsection{Proof of theorem 1}
\vspace{-2mm}
\begin{theorem}
Given $A\in R^{d\times d}$ a real square matrix. The following statements are equivalent:
\begin{enumerate}
\item $tr(A^k) = 0$ for $k = 1,\dots d$, where $tr(\cdot)$ is the trace of a matrix.
\item $\sigma(A) = \{0\}$, where $\sigma(A)$ is the set of eigenvalues of $A$.\\
\vspace{-4mm}
\end{enumerate}
\end{theorem}

\begin{proof}
Assume $A$ admits a Jordan form $A = PJP^{-1}$, where $J$ has eigenvalues of $A$ on the diagonal and $1$ or $0$ on the super diagonal and $P$ is an invertible matrix. Then $A^k = PJ^kP^{-1}$ and $tr(A^k) = tr(PJ^kP^{-1}) = \sum_{i=1}^{M} m_i \lambda^k_i $ where $m_i$ are $M$ positive integers that are the algebraic multiplicity of an corresponding unique eigenvalue $\lambda_i$. Clearly $M\leq d$.\\
(1) $\Rightarrow$ (2): 
Given $\sum_i m_i \lambda^k_i = tr(A^k) = 0$  $k = 1,\cdots M$ we have the linear system
\[\sum_i m_i \lambda^k_i  = \sum_i (\lambda_i m_i) \lambda^{k-1}_i= 0 \textup{ for } n =1,\dots M-1,\]
or equivalently in the matrix form $V_M [\lambda_1m_1,\;\dots,\; \lambda_M m_M ]^T = 0$, where $V_M$ is a Vandermonde matrix formed by eigenvalues $\lambda_i$, i.e., the elements of $V_M$ are $(V_M)_{ij} = (\lambda_j^{i-1})$. Since $V_M$ is invertible, we have $\lambda_i m_i = 0$, i.e., all nonzero eigenvaules are of algebraic multiplicity zero. This concludes that the only eigenvalue with non-zero algebraic multiplicity is $0$. \\
(2) $\Rightarrow$ (1): $tr(A^k) = tr(PJ^kP^{-1}) = \sum_{i=1}^{M} m_i \lambda^k_i  = \sum_{i=1}^{M} m_i (0)^i = 0$ 
 
\end{proof}
The spectral radius $\rho(A)$ provides an upper bound of the penalty term $tr(\exp(A))$ used in NO-TEARS
algorithm. To see this, using Jordan canonical decomposition, one can show that $tr(\exp(A)) = \sum_{\lambda_i \in \sigma(A)} e^{\lambda_i} \leq \sum_{\lambda_i \in \sigma(A)} e^{|\lambda_i|}  \leq d \cdot e^{\rho(A)}$. Hence, finding a nilpotent matrix with $\rho(A) = 0$ enforces $tr(\exp(A)) = d$, the condition for DAG used in NO-TEARS algorithm.
\vspace{-2mm}
\subsection{Gradient of the spectral radius}
In section \ref{sec:nobear_diff}, we propose to minimize a perturbed problem. The primary reason of adding a small constant to $W\odot W$ is to make the the objective fucntion differentiable. It is known that, if an eigenvalue $\lambda(A)$ of a matrix $A$ is simple, the gradient of $\lambda$ with respect to $A$ is $\frac{\partial \lambda }{\partial A_{i,j}} = \frac{v_iu_j}{u^Tv},$ 
where $u$ and $v$ are the left and right eigenvectors associatedd with the eigenvalue $\lambda$. Furthermore, when $A$ is strictly positive, its largest eigenvalue is guaranteed to be simple by the Perron-Frobenius theorem\cite{meyer2000matrix}. Adding a small constant $\epsilon$ to the nonnegative matrix $W\odot W$ can make the function $\rho(W\odot W + \epsilon)$ strictly positive and therefore differentiable with a well-defined gradient, which can be computed as 
$\frac{\partial \rho(W\odot W + \epsilon)}{\partial W_{i,j}} = 2\frac{W_{ij}v_iu_j}{u^Tv},$ where $u$ and $v$ are left and right eigenvectors of $W\odot W + \epsilon$. 
\vspace{-2mm}
\bibliographystyle{ws-procs11x85}
\bibliography{main}

\end{document}